\documentclass[twocolumn,aps,showpacs,preprintnumbers,superscriptaddress]{revtex4}
\usepackage{graphicx}
\usepackage{dcolumn}
\usepackage{bm}
\usepackage{amsmath,amssymb}
\begin{document}
\title{Quantum Phase Transitions of Light in the Dicke-Bose-Hubbard model}

\author{Soi-Chan Lei}
\affiliation{Physics Department, National Tsing-Hua University, Hsinchu 300, Taiwan}
\author{Ray-Kuang Lee}
\affiliation{Physics Department, National Tsing-Hua University, Hsinchu 300, Taiwan}
\affiliation{Institute of Photonics Technologies, National Tsing-Hua University, Hsinchu 300, Taiwan}
\date{\today}

\begin{abstract}
We extend the idea of quantum phase transitions of light in atom-photon system with Dicke-Bose-Hubbard model for arbitrary number of two-level atoms.
The formulations of eigenenergies, effective Rabi frequencies, and critical chemical potentials for two atoms are derived. With a self-consistent method, we obtain a complete phase diagram for two two-level atoms on resonance, which indicates the transition from Mott insulator to superfluidity and with a mean excitations diagram for confirmation.
We illustrate the generality of the method by constructing the dressed-state basis for arbitrary number of two-level atoms.
In addition, we show that the Mott insulator lobes in the phase diagrams will smash out with the increase of atom numbers. The results of this work provide a step for studying the effects with combinations of Dicke-like and Hubbard-like models to simulate strongly correlated electron systems using photons.
\end{abstract}
\maketitle
\date{\today}
\section{Introduction}
Quantum phase transition (QPT) is the phase transition that can only be accessed at absolute zero temperature, by the change of an external parameter or a coupling constant driven by quantum fluctuations \cite{Sachdev99}.
QPT has attracted intensive studies in interacting many-body problems, originally for strongly correlated electronic systems in condensed matter physics \cite{Chaikin}, and more recently for weakly interacting ultra-cold atomic system \cite{Greiner02}.
Typically, it is difficult to control and probe such exotic quantum phenomena in strongly correlated systems of electrons.
Optical lattices, artificial man-made crystals by interfering laser beams, offer a versatile platform for studying  trapped Bose gases.
In this situation Bose-Hubbard model including on site two atoms interaction and hopping between adjacent sites is used for describing the many-body dynamics from a Mott insulating to a superfluid phase in a gas of ultracold atoms with periodic potentials \cite{Jaksch98}.

Instead photons are non-interacting Bosons, and there is no possibility to have any quantum phase transitions in purely photon systems.
For pure Bose system, the conducting phase at zero temperature is presumably always superfluid \cite{Fisher89}.
However, engineered composites of optical cavities, few-level atoms, and laser light can form a strongly interacting many body system to study the concepts and methods in condensed matter physics from viewpoint of quantum optics.
In this case, photonic condensed-matter analogue is possible realized with state-of-art photonic crystals embedded with high-Q defect cavities.
Therefore, photons interacting with atoms should be much easier to study and probe the critical quantum phenomena such as QPTs in conventional condensed matter systems \cite {Illuminati06}.
The simplest light-atom system is photons interacting with a single two-level atom (TLA), described by the Jaynes-Cummings model \cite{Scullybook, Yamamoto99}. 
With an array of high-Q electromagnetic cavities each containing a single TLA in the photon-blockade regime, quantum phase transitions of photonic insulator (excitations localization) to superfluid (excitations delocalization) are predicted by the Bose-Hubbard model \cite{Greentree06, Hartmann06} and the XY spin model \cite{Angelakis06}.

As the number of TLAs increases, collective effects due to interactions of atoms among themselves give rise to intriguing many-body phenomena \cite{Yamamoto99}.
In quantum optics Dicke model describes the collective spontaneous emission of an initially excited ensemble of $N$ TLAs interacting with a common photon field \cite{Dicke53}, and has triggered numerous investigations of various physical effects.
A collection of atoms prepared in a certain initial state could decay collectively like a huge dipole with the emission of radiation not proportionally to the atom number $N$ but to $N^2$, with a phase transition between a normal to a {\it superradiant} state \cite{Narducci90}. 
Actually, when the maximal distance between any two of TLAs is much less than a typical wavelength, the coupling interaction for the photon field no longer depends on the individual coordinate of the atoms but on the collective pseudo-spin coordinate. 
One has to add these pseudo-spins of $N$ TLAs up to a single large pseudo-spin which is described by the collective angular momentum operators \cite {Yamamoto99}. 

Dicke model itself can offer the possible insights in the nature of QPT. 
With a size-consistent Hamiltonian for the Dicke model, squeezing of the photon field carries signatures of the associated quantum critical phenomena \cite{Jarrett06}.
QPTs of the Dicke-like and Bose-Hubbard-like models are both well-studied in optical and condensed matter systems, but the combination of these two models are never studied.
In this work, we extend the idea of QPTs of light in atom-photon system, proposed by A. D. Greentree {\it et al.} \cite{Greentree06}, from a single TLA interaction system described by the Jaynes-Cummings model to arbitrary number of TLAs by the Dicke model.
The problem we address here is $N$ identical TLAs which couple to a single mode quantized radiation field within ideal photon cavities in the photon blockade regimes.
With a self-consistent method, we numerically demonstrate that Mott-insulator to superfluid quantum phase transitions exist even in the Dicke model for arbitrary number of two-level atoms interacting with photons.
Detailed calculations of system eigenenergies, effective Rabi frequencies, and the critical chemical potentials for two TLAs are derived.
The results of this work provide a more general picture for simulating strongly correlated electron systems using photons.

This work is organized as follows, in Section \ref{s-model}, we describe the Dicke-Bose-Hubbard model used for studying QPT of light.
The eigensystem solutions based on the dressed-state bases for two atoms are derived in Section \ref{s-eig}.
Results of mean field phase diagram, average excitations, and the extension to arbitrary number of TLAs are given in Section \ref{s-result}.
Section \ref{s-final} gives the conclusion.
\section{Dicke-Bose-Hubbard Model}
\label{s-model}
Followed by the proposal by Greentree {\it et al.} \cite{Greentree06}, Bose-Hubbard model can be extended to realize the Mott insulator to superfluid phase transitions in 2D photonic bandgap cavity network by including atom-photon interaction.
The Hamiltonian for our extended Dicke-Bose-Hubbard model is given by combining photon hopping between identical cavities in the photon-blockade regimes and the repulsive on-site TLAs interaction with the presence of one-site energy chemical potentials,
\begin{eqnarray}
\label{eqHtol}
&& H=\sum_i{H^{DM}_i}-\kappa{\sum_{ij}{a^+_i}{a_j}}-\mu{\sum_i{N_i}},
\end{eqnarray}
where $i, j$ are the index for the individual cavities and range over all sites, $N_i$ is the total number of atomic and photonic excitations.
The second and third terms in the Hamiltonian, Eq.(\ref{eqHtol}), are Bose-Hubbard-like Hamiltonian.
The conserved particles in our system is the on-site total excitations $N_i= {a^+_i} {a_i} + {J^+_i} {J^-_i} $.
This conserved quantity are not pure photons, but the dressed photons which are mixtures of atoms and photons.
We have assumed that all the inter-cavity hopping energy of photons $\kappa_i \equiv \kappa$, and the chemical potential in the grand canonical ensemble $\mu_i \equiv \mu$ have no difference between cavities. 
The implementation of such photonic condensed-matter analogue is original proposed in Ref.\cite{Greentree06}.
The first term in Eq.(\ref{eqHtol}) is the onset Hamiltonian for $N$ TLAs interacting with a single mode field within a photon cavity, given by the Dicke model, i.e.
\begin{eqnarray}
\label{eqHDM}
H^{DM}_i=\varepsilon {J^+_i}{J^-_i}+{\omega}{a^+_i}{a_i}+\beta({a_i}{J^+_i}+{a^+_i}{J^-_i}),
\end{eqnarray}
where $\varepsilon$ is the transition energy for the TLAs, $\omega$ is the radiation field frequency, ${a^+_i} $ and ${a_i} $ are the photon creation and annihilation operators, and $J^+_i=\sum_j {\sigma^+_j}, J^-_i=\sum_j {\sigma^-_j}$ are the collective raising and lowering angular momentum operators, respectively.
The cavity mediated atom-photon coupling energy $\beta$ is assumed to be real here.
A superfluid order parameter $\psi$, to be physically real (i.e. $\psi^*=\psi$), has been introduced for the studying of the QPTs in our Dicke-Bose-Hubbard Hamiltonian.
The system is in the superfluid phase for a non-zero order parameter $\psi \neq 0$, while in the insulator phase for a zero order parameter $\psi = 0$.
With the mean field assumption $\langle a^+_i \rangle =\psi ^*$, we take decoupling approximation to investigate our Dicke-Bose-Hubbard Hamiltonian, i.e. ${a^+_i} {a_j} = {\psi^*}{a_j}+\psi{a^+_i}-|\psi|^2$ which is proportional to $\psi$.
Then the on-site mean field Hamiltonian of Eq.(\ref{eqHtol}) reads
\begin{eqnarray}
\label{eqmf} 
H^{MF}_i=H^{DM}_i-\kappa{\psi(a_i+a^+_i)}+\kappa{|\psi|^2}-\mu{N_i}.
\end{eqnarray}
This mean field Hamiltonian is the same for every site.
Without bothering the number of nearest neighbors around each idea photon cavity, we select three nearest neighbors per cavity in our simulation for the reason that it dose not affect our numerical results actually.
The interaction part $\beta ({a_i} {J^+_i} + {a^+_i} {J^-_i}) $ in Eq.(\ref{eqHDM}) actually does not change the field energy and commutes with arbitrary functions of the photon-number operator.
Hence, we choose the eigenstates of the total excitations $N_i$ to be the bare-states for the Dicke-Bose-Hubbard Hamiltonian.
Consequently the subspace of $N_i$ excitations spanned by $N_i + 1$ vectors is given by the direct product of atom and field states, i.e. ${|atom\rangle} {|photon\rangle} $ \cite{Buzck05, Ru-Fen07}.

In the following, we choose two TLAs as an example, $N=2$, for a clear illustration.
The extension of the Dicke-Bose-Hubbard Hamiltonian for arbitrary number of TLAs is given in the Section \ref{s-result}, which can be easily calculated by the same approach.
For two TLAs, the bare-states are ${|0,e^{\otimes {2}}\rangle}{|n-2\rangle}$, ${|g,e\rangle}{|n-1\rangle}$, and ${|g^{\otimes {2}},0\rangle}{|n\rangle}$ with photon number $n$ which runs from $0,1,2,3$ to $\infty$.
In our notation the collective angular momentum eigenstates that describe the two TLAs are denoted as ${|0,e^{\otimes {2}}\rangle}$ for the case all the two atoms are in the excited state, ${|g,e\rangle}$ for the case only one atom is in the excited state, and ${|g^{\otimes {2}},0\rangle}$ for the case all the two atoms are in the ground state.
These three bare-states are the normalized symmetric eigenstates of the noninteracting part $\varepsilon {J^+_i} {J^-_i} + {\omega}{a^+_i}{a_i}$ in Eq.(\ref{eqHDM}).
We neglect the dipole-dipole interaction, ${J^+_i} {J^-_j}, i\neq j$, and represent the two TLAs interacting with the same cavity field simultaneously in an idea photon cavity.
For arbitrary excitations, in order to calculate the transition amplitude of Eq.(\ref{eqmf}), we use these three complete symmetry degenerate bare-states replete with photons as the bases, i.e.
\begin{eqnarray}
&& {|0,e^{\otimes {2}}\rangle}{|0\rangle}, {|g,e\rangle}{|1\rangle}, {|g^{\otimes {2}},0\rangle}{|2\rangle},  \nonumber\\
&& {|0,e^{\otimes {2}}\rangle}{|1\rangle}, {|g,e\rangle}{|2\rangle}, {|g^{\otimes {2}},0\rangle}{|3\rangle}, \nonumber\\ 
&& \qquad\qquad \cdot \cdot \cdot \nonumber\\
&& {|0,e^{\otimes {2}}\rangle}{|k-2\rangle}, {|g,e\rangle}{|k-1\rangle}, {|g^{\otimes {2}},0\rangle}{|k\rangle}, \nonumber\\
&& \qquad\qquad \cdot \cdot \cdot \nonumber\\
&& {|0,e^{\otimes {2}}\rangle}{|n-2\rangle}, {|g,e\rangle}{|n-1\rangle}, {|g^{\otimes {2}},0\rangle}{|n\rangle}. \nonumber
\end{eqnarray}
Here totally $3n$ bare-state bases form a group for the whole Hilbert space.
Based on these on-site bases, we construct a $3n\times 3n$ transition amplitude matrix for Eq.(\ref{eqmf}), i.e.
{\scriptsize
\begin{widetext}
\begin{eqnarray}
\label{eqmat}
H^{MF} &=& \left[
\begin{array}{cccc}
\begin{array}{|ccc|} \hline {2}{\varepsilon}-{2}{\mu} &\sqrt{2}{\beta} &{0} \\
\sqrt{2}{\beta} &{2}{\varepsilon}+\omega-{3}{\mu} &\sqrt{4}{\beta} \\
{0} &\sqrt{4}{\beta} &{2}{\omega }-{2}{\mu}\\ \hline
\end{array} &
\begin{array}{ccc} {-}{\kappa}{\psi} & & \\ 
0 &{-}{\sqrt{2}}{\kappa}{\psi} & \\
0 &{0} &{-}{\sqrt{3}}{\kappa}{\psi}
\end{array} & & \\
\begin{array}{ccc} {-}{\kappa}{\psi} &{0} &{0}\\
&{-}{\sqrt{2}}{\kappa}{\psi} &{0}\\
& &{-}{\sqrt{3}}{\kappa}{\psi}
\end{array} &
\begin{array}{|ccc|} \hline {2}{\varepsilon}+\omega-{3}{\mu} &\sqrt{4}{\beta} &{0}\\
\sqrt{4}{\beta} &{2}{\varepsilon}+{2}{\omega}-{4}{\mu} &\sqrt{6}{\beta} \\
{0} &\sqrt{6}{\beta} &{3}{\omega}-{3}{\mu}\\ \hline
\end{array} & 
\ddots & \\
& 
\begin{array}{ccc}
{-}{\sqrt{2}}{\kappa}{\psi} &{0} & \\
& & \\
& & \\
\end{array} &
\ddots &
\begin{array}{|ccc|} \hline {2}{\varepsilon}+{(n-2)}{\omega}-{n}{\mu} &\sqrt{2(n-1)}{\beta} &{0}\\
\sqrt{2(n-1)}{\beta} &{2}{\varepsilon}+{(n-1)}{\omega}-{(n+1)}{\mu} &\sqrt{n}{\beta}\\
0 &\sqrt{n}{\beta} &{n}{\omega}-{n}{\mu} \\ \hline
\end{array}
\end{array}
\right]\nonumber\\
&&{+}{\kappa}{|\psi|^2}.
\end{eqnarray}
\end{widetext}
}Eq.(\ref{eqmat}) is the starting matrix elements for our Dicke-Bose-Hubbard Hamiltonian with mean field assumption.

\section{EigenSystem Solutions}
\label{s-eig}
\begin{figure}
\includegraphics[width=3.5in]{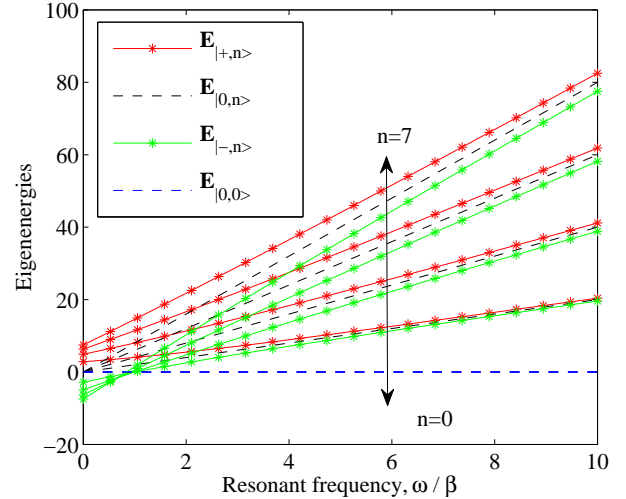}
\caption{Eigenspectrum for two TLAs in the Dicke-Bose-Hubbard model, as a function of the normalized resonant frequency $\omega/\beta$ for different photon numbers.
The eigenspectrum splits naturally into three branches, which are shown as positive $E_{|+,n\rangle}$ (upper), center $E_{|0,n\rangle}$ (centered), and negative $E_{|-,n\rangle}$ (lower) with the corresponding dressed-states, ${|+,n\rangle}$,  ${|0,n\rangle}$, and ${|-,n\rangle}$,respectively.}
\label{fEg}
\end{figure}

In general it is very difficult to diagonalize Eq.(\ref{eqmat}) and find out all the desired dressed-states even for two TLAs.
However, it may be instructive to diagonalize part of the Hamiltonian by assuming $\psi=0$ for the mean field Hamiltonian.
And a self-consistent method is applied for the case of $\psi \neq 0$ numerically.
In Eq.(\ref{eqmat}), those entries in the boxes correspond to the block diagonal form of Eq.(\ref{eqHDM}) with $\mu=0$ therein, and the numbers of excitations in each block increases by one.
As a matter of fact, we use the entries of the last box to calculate the corresponding eigenenergies and eigenstates.
Nevertheless, the analytic eigensystem solutions for non-zero detuning are very complicated, and we now focus on the case of zero detuning.
In the follows we assume the system is on resonance, i.e. the atomic frequency is the same as the field frequency $\varepsilon = \omega$.
In an idea photon cavity, when the two TLAs are excited, the rest of the photons will be dressed by the two TLAs.
The interaction, last term in Eq.(\ref{eqHDM}), couples the three bare-states in the same excitation $n$ manifold.
The eigenstates and eigenvalues are derived in Eq.(\ref{eqdr1}-\ref{eqdr4}) for the center ($E_{|0,n\rangle}$, $|0,n\rangle$) and upper/lower ($E_{|\pm,n\rangle}$, $|\pm,n\rangle$) branches.
The eigenspectrum splits naturally into three branches, corresponding to the upper branch $E_ {|+, n\rangle}$, centre branch $E_ {|0, n\rangle} $, and the lower branch $E_ {|-, n\rangle}$.
\begin{widetext}
\begin{eqnarray}
\label{eqdr1}
&& E_{|0,n\rangle}={n}\frac{\omega}{\beta},\\
\label{eqdr2}
&&|0,n\rangle=\frac{{-}{\sqrt{n-1}}{|0,e^{\otimes{2}}\rangle}{|n-2\rangle}+{\sqrt{n}}{|g^{\otimes{2}},0\rangle}{|n\rangle}}{\sqrt{2n-1}},\\
\label{eqdr3}
&& E_{|\pm,n\rangle}=\frac{(2n+1){\frac{\omega}{\beta}}{\pm}{R(n,\frac{\omega}{\beta})}}{2},\\
\label{eqdr4}
&&|\pm,n\rangle= \frac{\sqrt{n}{|0,e^{\otimes{2}}\rangle}{|n-2\rangle} + \frac{1}{2\sqrt{2}}[\frac{\omega}{\beta}{\pm}{R(n,\frac{\omega}{\beta})}]{|g,e\rangle}{|n-1\rangle}+ \sqrt{n-1}{|g^{\otimes{2}},0\rangle}{|n\rangle}}{\sqrt{2n-1+\{\frac{1}{2\sqrt{2}}[\frac{\omega}{\beta}{\pm}{R(n,\frac{\omega}{\beta})}]\}^2}},
\end{eqnarray}
\end{widetext}
Fig.\ref{fEg} shows the eigenenergies spectrum for two TLAs with the extended Dicke-Bose-Hubbard Hamiltonian.
The interaction part, $\beta ({a_i} {J^+_i} + {a^+_i} {J^-_i})$, in Eq.(\ref{eqHDM}) leads the $3n$ bare-states to couple together and form the normalized dressed-states.
These three dressed-states for the $n=1$ excitation manifold in Eqs.(\ref{eqdr1}-\ref{eqdr4}) can be reduced to the so called triplet states and the isolated singlet state, $|0, 0\rangle$ \cite{Brandes05}.
Our definition for the dressed-states extend to $n = 0$ is shown in Fig.\ref{fEg}, and we define the ground state for the dressed-state system as $|0,0\rangle$ with $E_{|0,0\rangle} = 0$.
With a non-trivial form of the raising operator, the ground state for the Dicke Hamiltonian is qualitatively different from other dressed-states in Eqs.(\ref{eqdr2}).
The branches emerge at large resonant frequency, and the splitting increases with larger photon number, $n$, given by the effective Rabi frequency.

\begin{figure}
\includegraphics[width=3.5in]{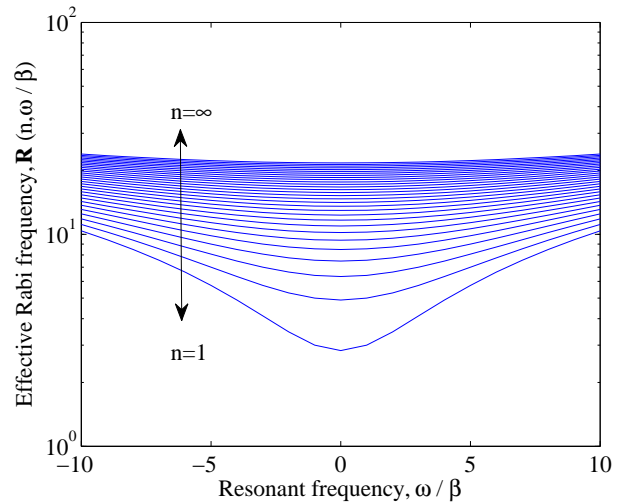}
\caption{Effective Rabi frequency $R(n,\omega/\beta)$ defined in Eq.(\ref{eqRabi}) v.s. normalized resonant frequency $\omega/\beta$.}
\label{fRabi}
\end{figure}

To have a clear picture of the photon-atoms interaction in our two TLAs system, we introduce an effective Rabi frequency as $R(n, \frac{\omega}{\beta})$ in Eq.(\ref{eqdr1}-\ref{eqdr4}), which for $n$  photons has the form, 
\begin{eqnarray}
R(n,\frac{\omega}{\beta})=\sqrt{8(2n-1)+(\frac{\omega}{\beta})^2}.
\label{eqRabi}
\end{eqnarray}
A two-state system has two possible states. i.e. a photon in our model can either be in the excited or ground state.
The effective Rabi frequency induced here can be used as a measurement for the energy splitting between the two states. 
For different photon numbers, the dressed-states described in Eqs.(\ref{eqdr1}-\ref{eqdr4}) oscillate with a different Rabi frequency which is not only proportional to the photon number of the field but also with a dependence on the dimensionless resonant frequency $\omega/\beta$ as well.
The dependence of the effective Rabi frequency on the photon number $n$ for different normalized resonant frequency $\omega/\beta$ is shown in Fig.\ref{fRabi}.
In the central region on the diagram, the boundary between domains defined by the effective Rabi frequency for different photon number eigenstate are the first state $|-, 1\rangle$, higher order states $|-, 2\rangle$ ,$|-, 3\rangle$, and so on.

With the dressed-state formalism in Eqs.(\ref{eqdr2}-\ref{eqdr4}), we can explain the features of Mott insulator to superfluid QPTs by a simple explanation \cite {Cirac91}.
Under the action of the Hamiltonian and the atomic population inversion operator, the whole Hilbert space splits into one- and two-dimensional subspaces with $n$ photons which are decoupled.
The probability amplitude achieves a saturate value when the number of photon increases.
These dressed states are time independent and with constant amplitude, therefore when the atom-field system is prepared in a dressed states the atom remains stationary \cite {Barnett97}.
Since we know these dressed-states and their corresponding eigenenergies in Eqs.(\ref{eqdr1}-\ref{eqdr4}), the corresponding wavefunction and the dynamics of the system will be just the superposition of them in addition to a phase term $\text{exp}(-iE_{|\{0,\pm\},n\rangle})t$.
In order to determine the ground state, we assume that $E_ {|-, n\rangle} < E_ {|+, n\rangle}$, i.e. the negative branch has lower eigenenergy.
The three degenerate bare-states for the $n$ excitation manifold form the three nondegenerate dressed-states at three energies, i.e. ${n{\frac{\omega}{\beta}}-\frac{1}{2}[{R(n,\frac{\omega}{\beta})}}-\frac{\omega}{\beta}]$, ${n}\frac{\omega}{\beta}$, and ${n{\frac{\omega}{\beta}}+\frac{1}{2}[{R(n,\frac{\omega}{\beta})}}-\frac{\omega}{\beta}]$.
The principal feature of the Rabi frequency spectrum in Fig. \ref{fRabi} is easily understood in terms of the partly dressed states ${|+, n\rangle}$ and ${|-, n\rangle}$ in Eq.\ref{eqdr4} which are separated in energy by an amount $\hbar{R(n,\frac{\omega}{\beta})}$.
On the other hand, the center branch appears to be a number of the two branch $E_ {|+, n-1\rangle}, E_ {|-, n\rangle}$ asymptotically.
\begin{figure}
\includegraphics[width=3.5in]{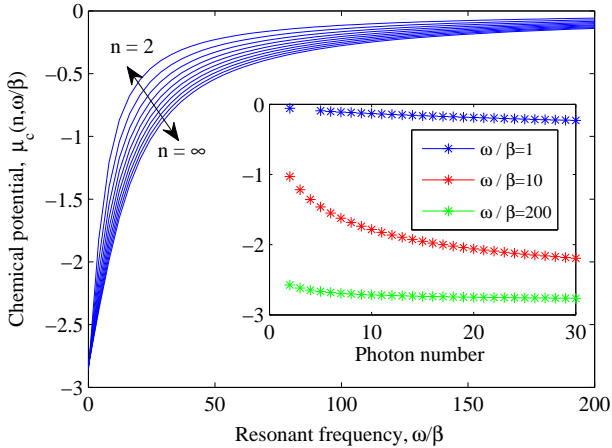}
\caption{The critical chemical potential $\mu_c{(n,\omega/\beta)}$ v.s. normalized resonant frequency $\omega/\beta$. The value of $\mu_c$ saturates as $\omega/\beta$ and $n$ increase in the limit with zero hopping $\kappa=0$.
Insert shows the dependence of the critical chemical potential for different photon numbers.}
\label{fmu}
\end{figure}

\begin{figure}
\includegraphics[width=3.5in]{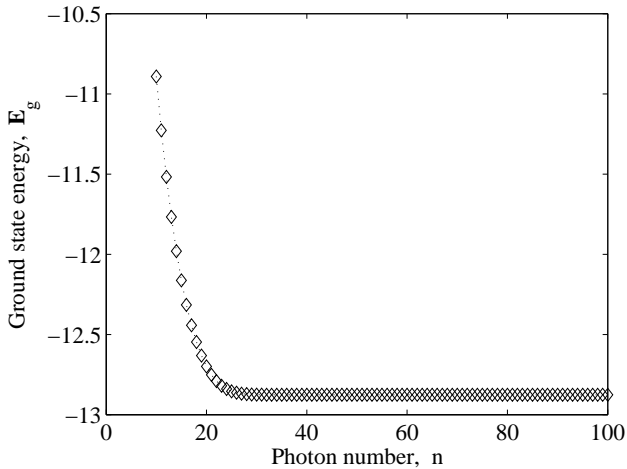}
\caption{The convergence of the ground state energy for different photon numbers.}
\label{fground}
\end{figure}

To derive analytical solutions for the dimensionless critical chemical potential, we substrate the negative branch energies, i.e. $E_ {|-, {n+1}\rangle}-E_ {|-, n\rangle}$, for that the system will change from $n$ to $n+1$ excitation per site and have the formula for different photon numbers, i.e.
\begin{eqnarray}
\label{equ}
&& \mu_c{(n,\frac{\omega}{\beta})}= \\
&&\frac{(\sqrt{n-1}-\sqrt{n}){\frac{\omega}{\beta}}-[\sqrt{n-1}{R(n+1,\frac{\omega}{\beta})}-\sqrt{n}{R(n,\frac{\omega}{\beta})}]}{2{\sqrt{2n(n-1)}}}.\nonumber
\end{eqnarray}
As one can see in Fig. \ref{fmu}, the critical chemical potential saturates to a constant value as the normalized resonant frequency and the photon number increase.
The dependence of the critical chemical potential with photon number is shown in the insert of Fig. \ref{fmu}, which is sensitive for the case of the normalized resonant frequency around $\omega/\beta\approx 10$.
This should be a suitable parameter for experimental observation \cite{Greentree06pra}.
In order to calculate the ground state wavefunction of our generalized Dicke-Bose-Hubbard Hamiltonian we numerically solve Eq.(\ref{eqmat}) by applying a self-consistent method.
When the photon number of the system increases, the ground state energy $E_g$ will converges to the true ground state of the system.
In Fig. \ref{fground}, we show the typical convergence of the ground state energy by increasing the photon numbers, which would quickly reach the minimum ground state energy for photon number up to $n = 30$.
When stimulated with light, the interaction of each cavity with an ensemble of these atoms gives rise to a composite optical-atomic state as in Eqs.(\ref{eqdr1}-\ref{eqdr4}).
However, if the photon number $n$ is smaller than or just equal to the atom number $N$, the quantum phase transition of light from Mott insulator to superfluid does not occur.
Instead the ground state of the Dicke model exhibits an infinite sequence of instabilities quantum-phase-like transitions\cite{Buzck05}.
\begin{figure}
\includegraphics[width=3.5in]{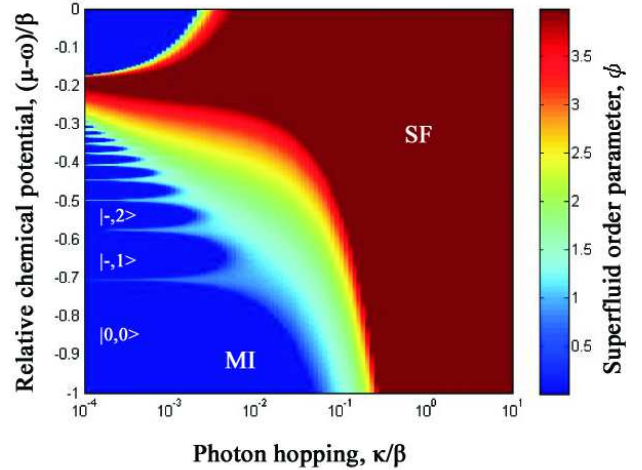}
\caption{Phase diagram of the mean field Hamiltonian for the ground state on resonance, shown by the normalized inter-cavity hopping energy of photons $\kappa/\beta$ and the relative chemical potential $(\mu-\omega)/\beta$.
The notation SF refers to a superfluid phase with strong interaction of photon hopping while MI refers to a Mott insulator phase with equally number of photons in each cavity.
In the insulator region $|0, 0\rangle$, $|-, 1\rangle$, and $|-, 2\rangle$ denote the negative branches of the dressed-states where the system will change from $n$ to $n+1$ excitation per site simultaneously filling photons in cavities and resulting in a finite gap of spectrum.}
\label{Fig:2}
\end{figure}
\begin{figure}
\includegraphics[width=8.15cm]{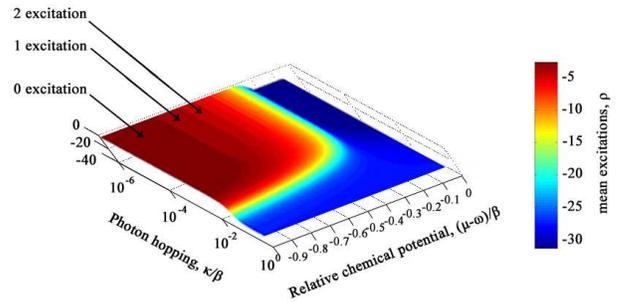}
\caption{The mean excitations for two TLAs. First few plateaus indicate constant density regions of the excitations $0$, $1$,and $2$ which correspond to the ground state configurations $|0,0\rangle$, $|-,1\rangle$, and $|-,2\rangle$, respectively.}
\label{Fig:3}
\end{figure}
\begin{figure}
\includegraphics[width=4.15cm]{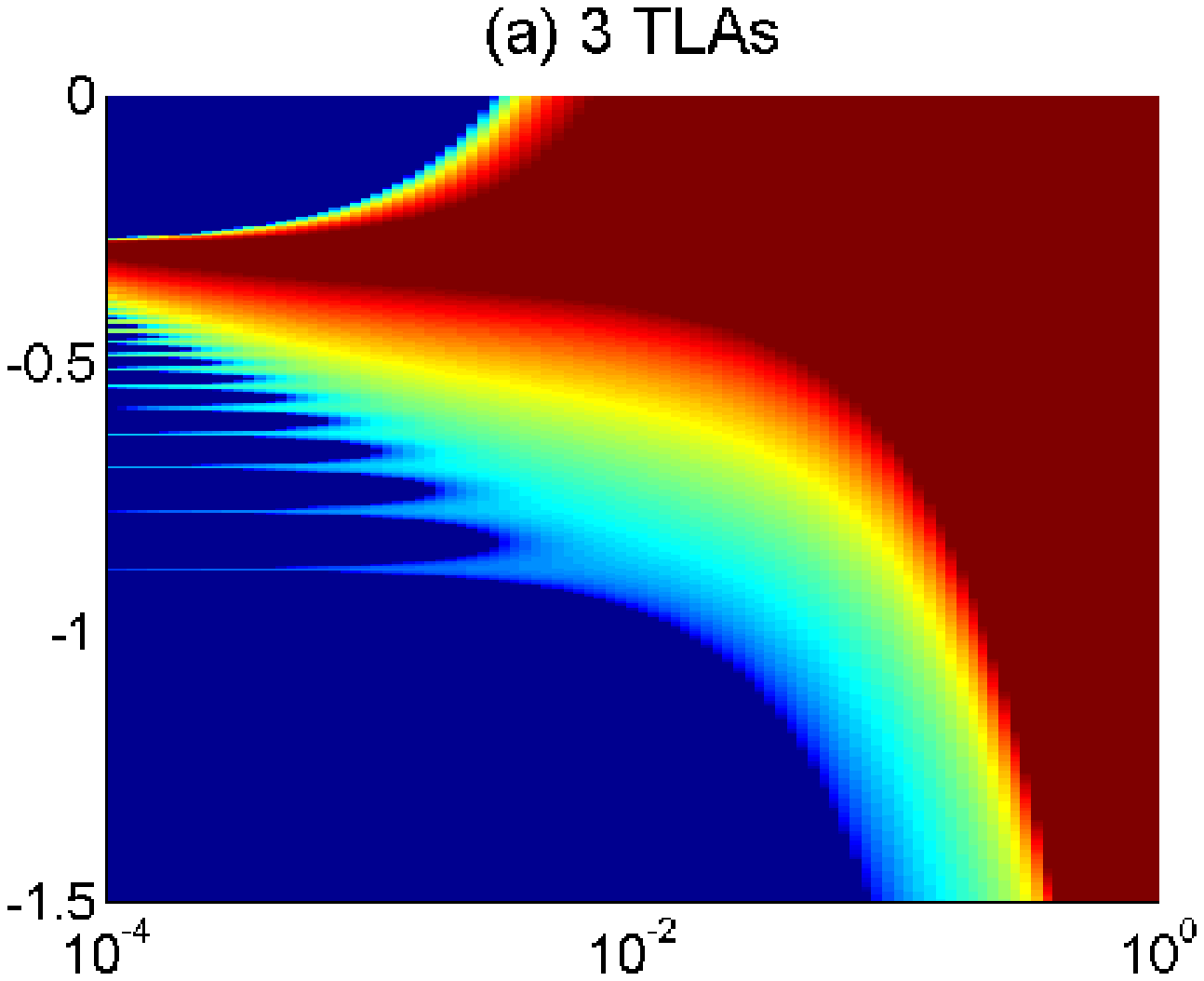}
\includegraphics[width=4.15cm]{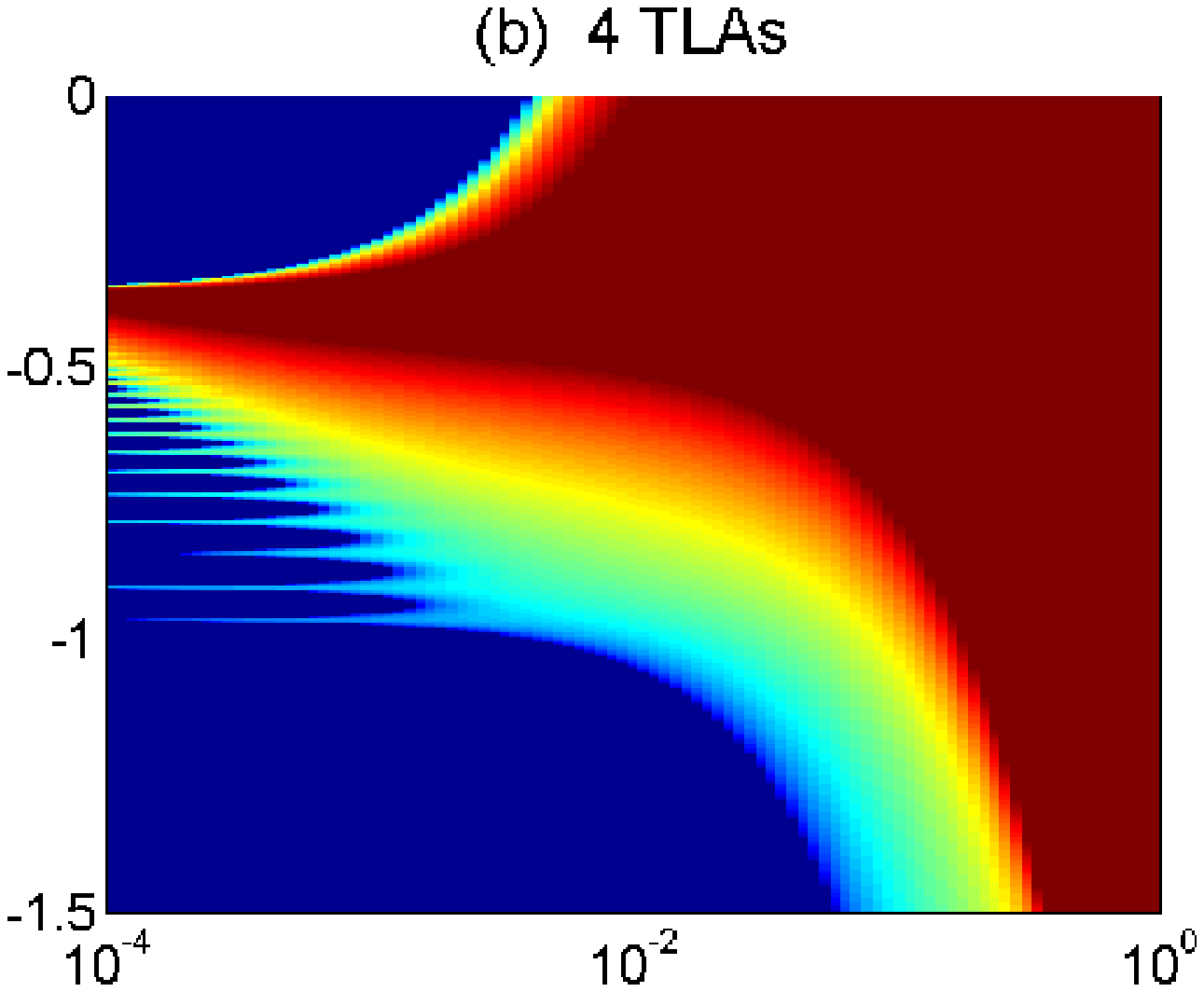}
\includegraphics[width=4.15cm]{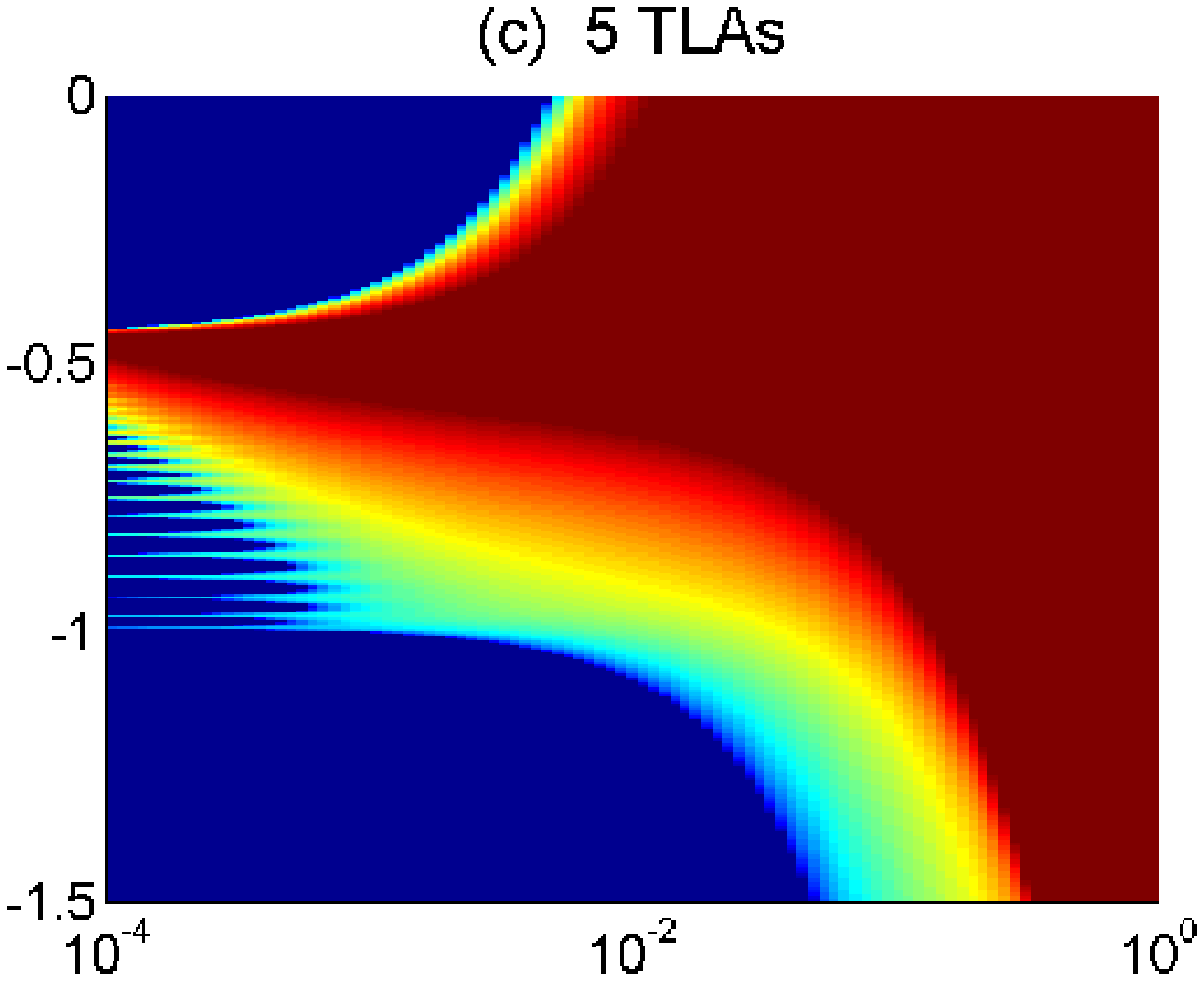}
\includegraphics[width=4.15cm]{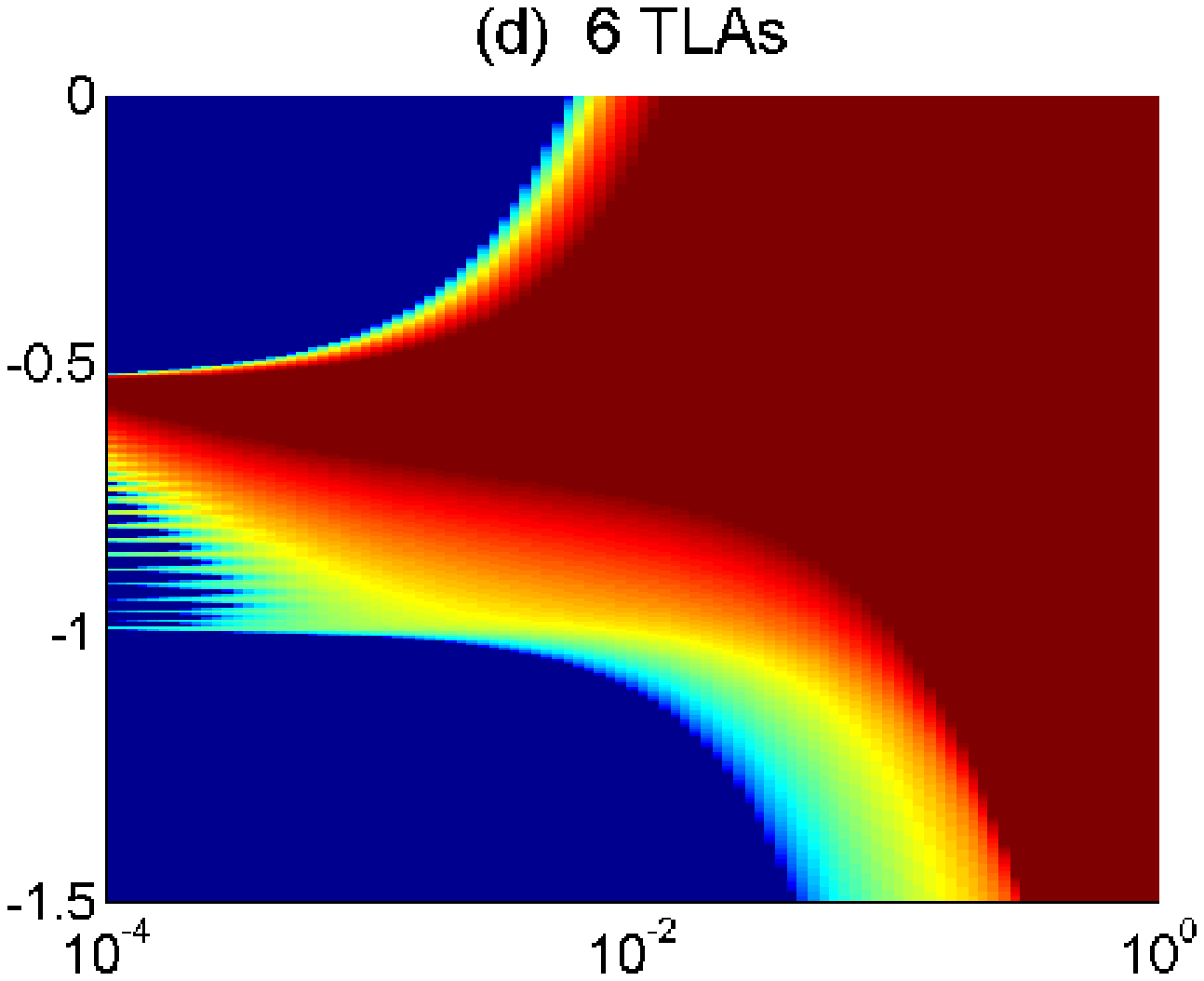}
\includegraphics[width=4.15cm]{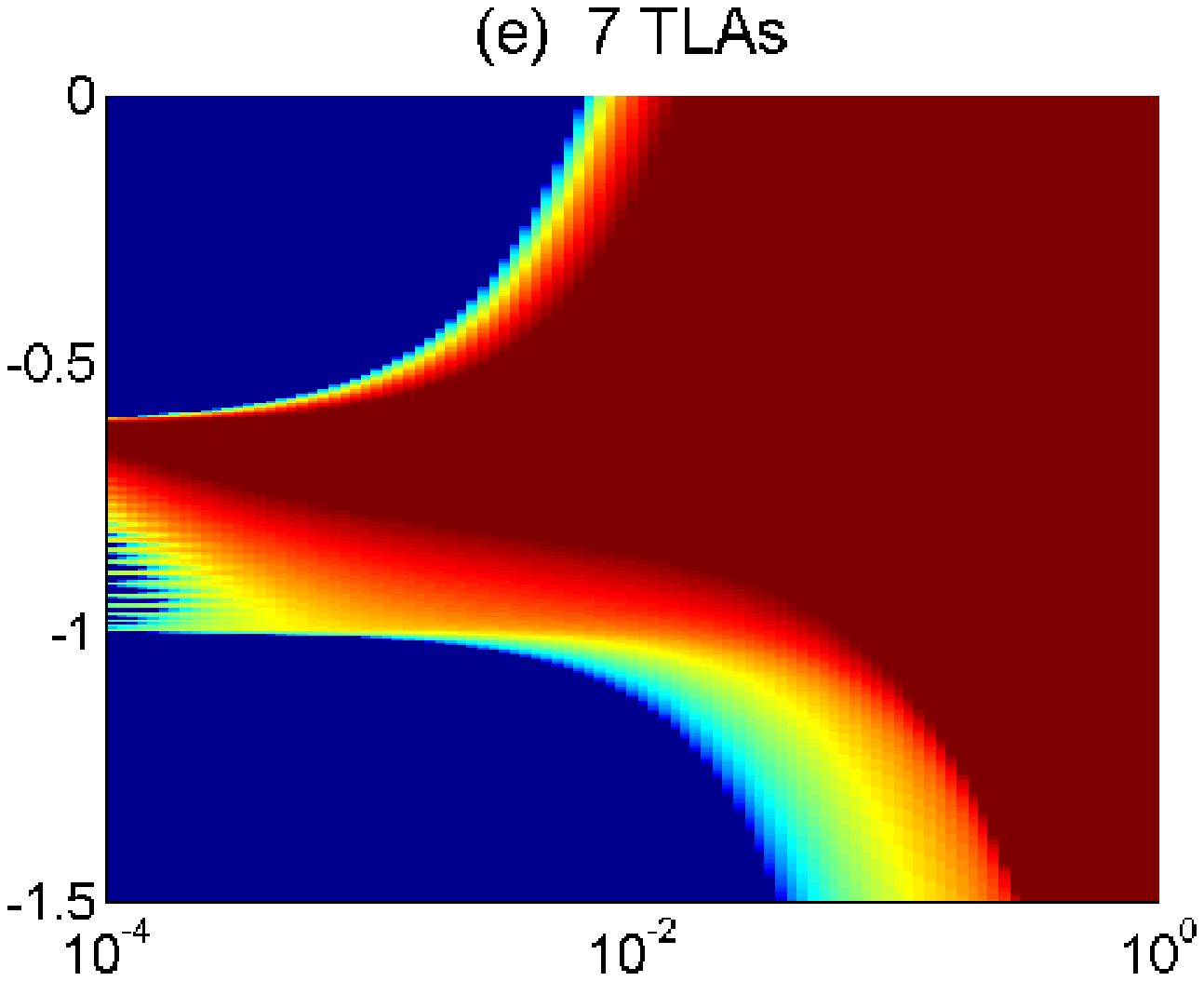}
\includegraphics[width=4.15cm]{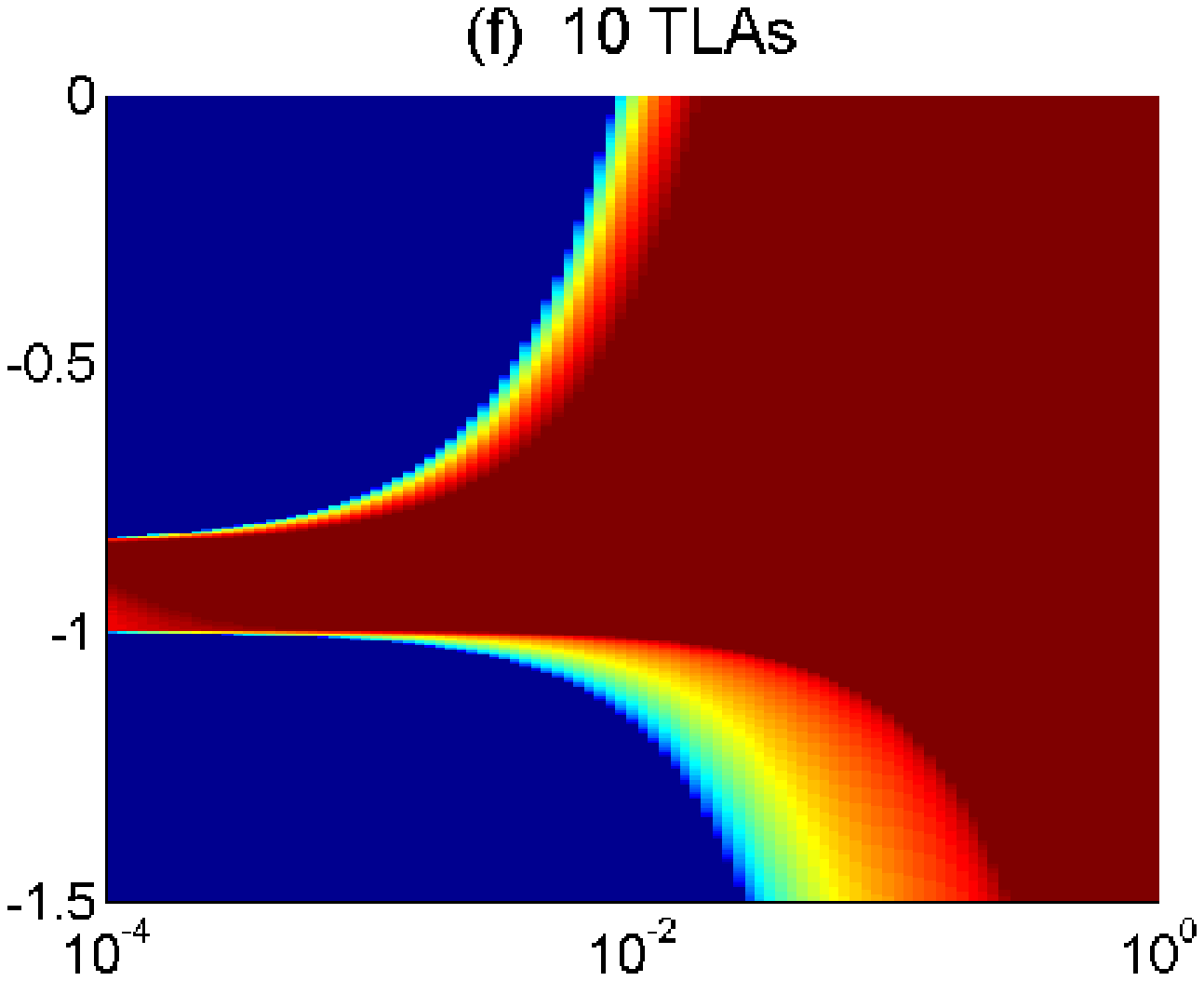}
\caption{The phase diagrams for arbitrary number of TLAs, (a) $N =3$, (b) $N =4$, (c) $N = 5$, (d) $N = 6$, (e) $N=7$, and (f) $N=10$.
The axises of horizontal and vertical are defined as the same as in Fig.\ref{Fig:2}.
The phase boundary between MI-SF are the superposition states that disappear and result in a single macroscopic coherent radiation state for large number of TLAS.}
\label{Fig:4}
\end{figure}
\section{Results}
\label{s-result}
\subsection{Mean field phase diagram}
After deriving the formulations for two TLAs in our system, we calculate the phase diagram for the Dicke-Bose-Hubbard Hamiltonian in Eq.(\ref{eqmf}) for two TLAs by applying the mean field theory and the self-consistent method, as shown in Fig.\ref{Fig:2}.
Clear quantum phase transitions can be seen in the diagram for different normalized inter-cavity hopping energy of photons $\kappa/\beta$ and different relative chemical potential $(\mu-\omega)/\beta$.
Here the notation SF refers to a superfluid phase with strong interaction of photon hopping.
And the notation MI refers to a Mott insulator phase with equally number of photons in each cavity.
The Mott insulator to superfluid phase transitions occur for the case of the photon number is larger than the atom number, $N \ll n$.
We give a simple picture for the QPTs of light in our system.
As photons pass through an array of high-Q cavities with two TLAs per site, there is an upper limit for the energy of the two TLAS as all the atoms have been excited to the upper level.
At the end, interaction photons are not independent to each other due to the iterations from effective on-site repulsion, and inter-cavity hopping with their proximity.
In such a way the two TLAs are strongly correlated.
Change of the critical chemical potential results in the change of the on-site numbers of photons, as shown in Fig.\ref{Fig:2}.
Regarding to the on-site chemical energy $\mu$, the regions to the right corresponding to chemical potential $\mu\neq 0$ \cite{Sachdev99}.
In the insulator region $|0, 0\rangle$, $|-, 1\rangle$, and $|-, 2\rangle$ denote the negative branches of the dressed-states where the system will change from $n$ to $n+1$ excitation per site, simultaneously filling photons in cavities and resulting in a finite gap of spectrum.
The superfluid phase is the eigenstates of $a_i$ and excitations over the ${|-, n\rangle} $ branches in the Fig.\ref{Fig:2}.
The probability of finding the average photon number $\overline {n} $ in this regime obeys Poisson distribution.
With increasing laser intensity $\kappa \ll \beta $, we have strong interactions.
It is the most important regime with rich dynamics where the on-site repulsion dominates with equally numbers of photons in each cavity as one can see on the left in Fig.\ref{Fig:2}.
This region is corresponding to a constant density of photons filling in cavities simultaneously.
In this situation each site has exactly the same integer number of photons with strongly on-site coupling regime.
A finite gap of spectrum is formed and photons here are incompressible, resulting an insulator phase on the other hand \cite{Fisher89, Greiner02, Stoferle04}.
\subsection{Average excitations}
Another way to indicate QPTs in our Dicke-Bose-Hubbard Hamiltonian is to study the average excitations.
We minimize the ground state energy with respect to the order parameter, $\psi$, and take the derivative of the ground state energy with respect to $\mu$.
Theoretically, we can consider the problem by fixing the chemical potential by varying photon numbers in the grand-canonical ensemble and the derivatives value can be expressed by the ensemble average, i.e.
\begin{eqnarray}
&& \rho= {- \frac{\partial E_g{(\kappa,\mu)}}{\partial \mu}}|_{\psi =\psi_{min}}.
\end{eqnarray}
The mean excitations for two TLAs in our systems are shown in Fig. \ref{Fig:3}.
It can be clearly seen that the density of the incompressible Mott insulator phases does not change with the change of the relative chemical potential, and the value of the relative chemical potential jumps discontinuously goes through a lobe.
Regions with varying $\rho$ have coherent states as ground state configurations.
In fact, we use the mean excitations to confirm the numbers of photons in each Mott lobe, and it is the mean excitation $\rho$.
On can view this average excitations for the evidences of the phase transitions and lobes manifest in Fig. \ref{Fig:2}. 
\subsection{Extension to arbitrary number of TLAs}
For arbitrary number of TLAs, $N$, and arbitrary number of photons, $n$, we use following general bases for Eq.(\ref{eqmf}) to solve our Dicke-Bose-Hubbard Hamiltonian,
\begin{widetext}
\begin{eqnarray}
&& {|0,e^{\otimes {N}}\rangle}{|0\rangle}, {|g^{\otimes({N-1})},e\rangle}{|1\rangle},\cdot \cdot \cdot \, {|g,e^{\otimes ({N-1})}\rangle}{|{N-1}\rangle},{|g^{\otimes {N}},0\rangle}{|{N}\rangle},  \nonumber\\
&& {|0,e^{\otimes {N}}\rangle}{|1\rangle}, {|g^{\otimes({N-1})},e\rangle}{|2\rangle},\cdot \cdot \cdot \, {|g,e^{\otimes ({N-1})}\rangle}{|{N}\rangle},{|g^{\otimes {N}},0\rangle}{|N+1\rangle},  \nonumber\\
&&\qquad \qquad \qquad \cdot \cdot \cdot \nonumber\\
&& {|0,e^{\otimes {N}}\rangle}{|k-N\rangle}, {|g^{\otimes({N-1})},e\rangle}{|k-N+1\rangle},\cdot \cdot \cdot \, {|g,e^{\otimes ({N-1})}\rangle}{|{k+N-1}\rangle},{|g^{\otimes {N}},0\rangle}{|k\rangle},  \nonumber\\
&&\qquad \qquad \qquad\cdot \cdot \cdot \nonumber\\
&& {|0,e^{\otimes {N}}\rangle}{|n-N\rangle}, {|g^{\otimes({N-1})},e\rangle}{|n-N+1\rangle},\cdot \cdot \cdot \, {|g,e^{\otimes ({N-1})}\rangle}{|{n+N-1}\rangle},{|g^{\otimes {N}},0\rangle}{|n\rangle}.  \nonumber
\end{eqnarray}
\end{widetext}
Strong coupling with many excitations is expected to be first observed for a small number of $N$ \cite{Laussy05}, there we provide numerical results for a comparison from smaller to larger value of atom numbers $N$.
By taking the limitation of an arbitrary numbers of TLAs with fixed photon numbers up to $30$, we obtain the Mott insulator to superfluid phase diagrams for the number of TLAs from $N = 3$ to  $N = 10$ in Fig.\ref{Fig:4}, respectively.
By increasing the number of TLAs with the same amount of deposited photons, resulting in a insulating to superfluid phase transition which is characterized by the MI-SF transitions of regular bosons on the regions above the first lobe.
The value of $\kappa$ at the tip of the $n$-th Mott lobe varies as $\thicksim\frac{n}{N}$ for large $N$.
There is no energy barrier to the addition of extra photons and superfluidity occurs at arbitrarily small $\kappa$ \cite{Fisher89}.
When the limits ${N} \rightarrow\infty$ or $\beta\rightarrow\infty$ or both are taken, one has a single macroscopic coherent radiation state and the superposition of dressed-states disappears \cite{Frasca04}.
We can thus conclude that as the number of TLAs increases, superposition states may be disappearing and classically emerges.
\section{Conclusion}
\label{s-final}
With the Dicke-Bose-Hubbard Hamiltonian, we show that the Mott insulator to superfluid quantum phase transitions with photons can be realized in an extended Dicke model for arbitrary number of two-level atoms.
We illustrate the generality of the method by constructing the dressed-state basis for arbitrary number of two-level atoms.
Moreover, we show that as the number of TLAs increases, superposition states may be disappearing and classically emerges.
With more controllable lightwave technologies, the understanding of quantum phases transitions of light with distinctive properties, organizations of the ground state wave function, and practicable new applications should make it more easily to be realized. 
With combinations of Dicke-like and Hubbard-like models to simulate strongly correlated electron systems using photons, we believe that there would be more and more interesting quantum phase transitions of light to be demonstrated as those in condensed matter physics.

The authors thank A. D. Greentree and Chaohong Lee for their useful discussions and Yu-Sheng Hung for his help on the plotting the figures.
This research is supported by the National Science Council of Taiwan, NSC 95-2112-M-007-058-MY3 and NSC 95-2120-M-001-006.

\end{document}